\documentclass[conference]{IEEEtran}
\usepackage{cite}
\usepackage{amsmath,amssymb,amsfonts}
\usepackage{algorithmic}
\usepackage{braket}
\usepackage{graphicx}
\usepackage{textcomp}
\usepackage[dvipsnames]{xcolor}
\usepackage{fancyhdr}
\usepackage{comment}
\usepackage[hyphens]{url}
\usepackage{url}

\usepackage{soul}
\usepackage{booktabs}
\usepackage[bookmarks=true,breaklinks=true,letterpaper=true,colorlinks,linkcolor=black,citecolor=blue,urlcolor=black]{hyperref}
\usepackage{adjustbox}

\usepackage[linewidth=1pt]{mdframed}

\usepackage{tikz}

\usepackage{mathtools}

\usepackage[font=bf]{caption}

\usepackage{tabularx}

\usepackage[caption=false]{subfig}

\newcommand{\subparagraph}{}
\usepackage[utf8x]{inputenc}
\usepackage{authblk}

\def\BibTeX{{\rm B\kern-.05em{\sc i\kern-.025em b}\kern-.08em
    T\kern-.1667em\lower.7ex\hbox{E}\kern-.125emX}}

\title{Summary: Chicago Quantum Exchange (CQE) Pulse-level Quantum Control Workshop}

\author{Kaitlin N. Smith$^1$, Gokul Subramanian Ravi$^1$, Thomas Alexander$^2$, Nicholas T. Bronn$^2$ \\  Andre Carvalho$^3$, Alba Cervera-Lierta$^4$,
Frederic T. Chong$^1$, Jerry M. Chow$^2$, Michael Cubeddu$^5$ \\Akel Hashim$^6$, Liang Jiang$^7$, Olivia Lanes$^2$, Matthew J. Otten$^8$,
David I. Schuster$^7$ \\ Pranav Gokhale$^9$, Nathan Earnest$^2$, Alexey Galda$^{10,11,12}$
\\
\textit{\small$^1$Department of Computer Science, University of Chicago}\\
\textit{\small$^2$IBM Quantum, IBM T.\ J.\ Watson Research Center}\\
\textit{\small$^3$Q-CTRL}\\
\textit{\small$^4$Department of Chemistry and Department of Computer Science, University of Toronto}\\
\textit{\small$^5$Aliro Technologies, Inc.}\\
\textit{\small$^6$Department of Physics, UC Berkeley}\\
\textit{\small$^7$Pritzker School of Molecular Engineering, University of Chicago}\\
\textit{\small$^8$HRL Laboratories, LLC} \\
\textit{\small$^9$Super.tech}\\
\textit{\small$^{10}$Computational Science Division, Argonne National Laboratory}\\
\textit{\small$^{11}$James Franck Institute, University of Chicago}\\
\textit{\small$^{12}$Menten AI, Inc.}\\
}

\begin{document}

\maketitle
\pagestyle{plain}


\begin{abstract}

Quantum information processing holds great promise for pushing beyond the current frontiers in computing. Specifically, quantum computation promises to accelerate the solving of certain problems, and there are many opportunities for innovation based on applications in chemistry, engineering, and finance. To harness the full potential of quantum computing, however, we must not only place emphasis on manufacturing better qubits, advancing our algorithms, and developing quantum software. To scale devices to the fault tolerant regime, we must refine device-level quantum control.
  
  On May 17-18, 2021, the Chicago Quantum Exchange (CQE) partnered with IBM Quantum and Super.tech to host the Pulse-level Quantum Control Workshop. At the workshop, representatives from academia, national labs, and industry addressed the importance of fine-tuning quantum processing at the physical layer. The purpose of this report is to summarize the topics of this meeting for the quantum community at large.

\end{abstract}

\section{Introduction}

\textbf{Quantum computing today:} The present era of quantum computing is characterized by the emergence of quantum computers (QCs) with dozens of qubits. Although these devices are not fault tolerant, new algorithms exist that have innate noise resilience and modest qubit requirements. 
There are promising indications that near-term devices could be used to accelerate or enable solutions to problems in domains ranging from molecular chemistry~\cite{kandala2017hardware} to combinatorial optimization~\cite{moll2018quantum} and machine learning~\cite{biamonte2017quantum}. 

\textbf{Why take a pulse approach:} The underlying evolution of a quantum system is continuous and so are the control signals.
These continuous control signals offer much richer and more flexible controllability than the gate-level quantum instruction set architecture (ISA). The control pulses can drive the QC hardware to the desired quantum states by varying a system-dependent and time-dependent quantity called the Hamiltonian. 
The Hamiltonian of a quantum system is an operator corresponding to the total energy of the system. Thus, the system Hamiltonian determines the evolution path of the quantum states. 

The ability to engineer the real-time system Hamiltonian allows us to navigate the quantum system to the quantum state of interest through generating accurate control signals. Quantum computation can be done by constructing a quantum system in which the system Hamiltonian evolves in a way that aligns with a QC task, producing the computational result with high probability upon final measurement of the qubits.

\textbf{Pulse-level challenges:} While the benefits to the pulse approach are clear, there are several challenges stemming from the inherent complexities of a full-stack approach, some of which are discussed below:

\begin{enumerate}
    \item \emph{Machine Hamiltonian}: Pulse-level optimization typically requires an extremely accurate model of the quantum system or machine, i.e. its Hamiltonian. Hamiltonians are difficult to measure experimentally and moreover, they drift significantly over time between daily recalibrations. Experimental quantum optimal control (QOC) papers incur considerable overhead associated with pre-execution calibration to address this issue.
    \item \emph{Programming Model}: A traditional gate-level quantum programming  model  is  simply  an  abstraction  of  the  real quantum  hardware  execution  in  a  form  which  is  amenable to  users  familiar  with  classical  programming  models.  Thus,  execution  of  a  program  represented  via  the  quantum  circuit  model requires  translating circuit  instructions  to  pulses  which  enact the  desired  state-transformations  and/or  measurements.  This translation is often sub-optimal due to the heavy abstractions imposed across the software stack, resulting in a pulse-level instruction that closely approximates the ideal gate within some margin of error. Masking pulses with gate-level programming prevents exposing too much information to the end user, simplifying computation. Low-level quantum programming, if not done in a systematic manner, can considerably increase the complexity of an algorithm's specification.
    \item \emph{Optimization and Compilation Overheads}: Generating optimal pulses is an arduous task. Compilation and optimization overheads are especially prohibitive in applications such as variational quantum algorithms wherein the circuit and pulse construction process are part of the critical execution loop.  This   could potentially amount to several weeks of total compilation latency over the course of  thousands  of  iterations. 
    \item \emph{Simulation}: Quantum systems are dynamic and evolve with time. As a result, simulation tools for quantum mechanical systems need to take variation into consideration, whether that variation stems from intentional  gate  application  or  from  unintentional  drift or  environmental  coupling. If quantum  systems are modeled  with  enough  granularity, pulses required for low-level control can be developed. 

\end{enumerate}

\textbf{Role of workshop and report:}

There are many challenges, as previously mentioned, associated with low-level quantum control. Overcoming these difficulties, however, could contribute to achieving quantum advantage in the near-term. Additionally, accelerated quantum processing could emerge as a significant benefit of custom architectures produced as a result of software-hardware co-design.

The Chicago Quantum Exchange (CQE) Pulse-level Quantum Control Workshop was organized with the intent of informing the broader community of the advantages of pulse-level quantum programming. Included sessions did not all specifically focus on pulse-level software or hardware optimization, but the role of low-level control in improving near-term QCs was an integral theme throughout the workshop. Talks featured many types of quantum technologies and information encodings, showing the potential for wide range of improved control in the hardware-diverse quantum space. 

The purpose of this report is to summarize the topics discussed during the workshop and to provide a reference for the quantum community at large.

\section{Quantum Information}
\label{quantum-information}

Quantum information science (QIS) redefines the classical computational model through the use of a type of information that can hold many values at once. Most frequently, radix-2 or base-2 quantum computation is implemented within algorithms and quantum computer architectures. This type of quantum computation uses quantum bits, or qubits, that have two basis states represented as  $\Ket{0} =  \begin{bmatrix}  1 & 0
 \end{bmatrix}^\text{T}$ and $\Ket{1} =  \begin{bmatrix}  0 & 1
 \end{bmatrix}^\text{T}$. Qubits, unlike classical bits that hold a static value of either 0 or 1, demonstrate states of superposition in the form of $\alpha_0\ket{0} + \alpha_1\ket{1}$ with probability amplitudes $\alpha_0,\alpha_1 \in \mathbb{C}$ such that $|\alpha_0|^2+|\alpha_1|^2=1$. Superposition enables $n$ qubits to represent states in $2^n$-dimensional Hilbert Space, and this phenomenon, along with the ability for quantum states to interfere and become entangled, allow certain problems to be solved with significant reductions in complexity. Qubits hold large quantities of information for processing while in superposition, but upon measurement, the quantum state collapses; only classical values of either 0 or 1 are observed. 
 
 Radix-$d$ computation where $d>2$ is seen occasionally in classical systems, particularly in domain-specific applications. However, the benefits of high-dimensional encoding are often outweighed by the advantages provided through the continuous scaling of bistable transistors. 
Similarly, higher-dimensional quantum computation
 is not infeasible and trade-offs associated with its application are actively being explored. The qudit, or quantum digit, is the multi-level quantum unit that can be used as an alternative to the base-2 qubit. A qudit is described by a $d$-dimensional vector and is written as
 
 \begin{equation}
     \Ket{\Psi} = \sum_{i=0}^{d-1} \alpha_i\Ket{i}
 \end{equation}

\noindent where $\alpha_i$ values are the probability amplitudes corresponding to the basis states $\Ket{i}$. As with the qubit, qudit probabilities must all sum to one.

Under the \textit{no-cloning theorem}, unknown qubit and qudit states cannot be copied without destroying superposition. In other words, any attempt to duplicate quantum information essentially acts as a measurement operation, resulting in a basis state. As a result, quantum error correction and information storage methods that preserve state cannot be implemented like their classical analogs because classical processing often exploits the ability to efficiently copy data. The no-cloning theorem inflicts a serious architectural constraint for quantum information processing.

Quantum computation is implemented with operators or gates that cause the probability amplitudes associated with each basis in the quantum state to evolve. These operators are represented by a unitary transformation matrix, $U$, of size $d^n \times d^n$ where $n$ is the number of radix-$d$ units of quantum information that the operation transforms. Quantum operations are reversible since gates are unitary. $UU^\dagger = U^\dagger U = I$ where the symbol $^\dagger$ indicates a complex conjugate operation that creates the inverse of $U$, and $I$ is the identity operation that preserves quantum state. Measurement is not reversible since it causes quantum states to collapse to classical information. 

Quantum operations are necessary for qubits and qudits to demonstrate special quantum properties. Single input operations can be used to create superpositions of basis states. If entangled states are desired, multi-qubit or multi-qudit gates must be available. When qubits or qudits are entangled, they act as an inseparable system where any action on one part of the system impacts the other(s).

Cascades of quantum operations create quantum circuits or algorithms. The set of gates employed by the circuit depends on the abstraction level used in the description. For example, higher-level circuits that are technology-independent may use complex, multi-qubit or qudit operators whereas lower-level circuits targeted for execution on quantum hardware will implement a set of elementary single- and two-qubit or qudit gates. The set of basis gates used for a specific QC is technology-dependent as certain quantum platforms implement some gates more efficiently than others. A basis gate set usually consists of a set of single-qubit or qudit operations that implement arbitrary rotations within a small margin of error along with a a multi-qubit or qudit operation, such as the radix-2 $CX$ or $CZ$ gates, to form a universal gate set.

\section{Quantum Algorithms}
\label{quantum circuits}

Search, simulation, optimization, and algebraic problems are all proposed for QCs, but the breadth of quantum applications will depend on the size and capability of available machines. The subset of computations, as well as their corresponding quantum kernels, for which QCs will promise an advantage is still being defined, and over time, it is likely that this class of problems will evolve. A key challenge for quantum researchers is developing efficient methods for categorizing problems for the best suited hardware, either classical or quantum, and then running subroutines derived from partitioned algorithms accordingly. 

\subsection{Variational Algorithms}
\label{variational algoithms}

Variational quantum algorithms (VQA) provide an exciting opportunity for near-term machines to demonstrate quantum advantage. VQAs have a one-to-one mapping between logical qubits in algorithms to physical qubits in QCs. Physical qubits will be discussed more in Section~\ref{Physical Qubits}. Proposed applications of VQAs include ground state energy estimation of molecules~\cite{peruzzo2014variational} and MAXCUT approximation~\cite{moll2018quantum}. VQAs are hybrid algorithms since they comprise of classical and quantum subroutines.

VQAs are well suited for near-term hardware since they adapt to the intrinsic noise properties of the QC they run on. The VQA circuit is parameterized by a vector of angles that correspond to gate rotations on qubits. The vector of angles is optimized by a classical optimizer during many iterations to either maximize or minimize an objective function that represents the problem that the VQA implementation hopes to solve. 

Two notable VQAs are the Variational Quantum Eigensolver (VQE) and the Quantum Approximate Optimization Algorithm (QAOA). The former is often used for quantum chemistry while the latter is applied to combinatorial optimization.

\subsection{Fault Tolerant Algorithms}
\label{fault tolerant algoithms}

Fault tolerant quantum algorithms offer exciting computational gains over their classical analogs, but they require extremely low error thresholds. Because of this, error correcting codes (ECCs) are required that enforce a one-to-many mapping between logical qubits in an algorithm to physical qubits on a QC. This encoding allows quantum information to be shielded from errors that arise due to uncontrolled coupling with the environment or due to imperfect operations at the physical level. There are many kinds of quantum ECCs~\cite{devitt2013quantum}. 

At minimum, it is estimated that millions of high-quality physical qubits will be needed to implement fault-tolerant quantum computation~\cite{o2017quantum}. Once fault tolerance is reached, disruptive implementations of Shor's algorithm for quantum factoring~\cite{shor1999polynomial}, Grover's quantum database search~\cite{grover1996fast}, and quantum phase estimation~\cite{mike_ike_2020} can be implemented.

\section{Quantum Hardware}
\label{Hardware}

\subsection{Physical Qubits}
\label{Physical Qubits}

Even if the theory is well defined, information must have a physical realization for computation to occur. Currently, a variety of quantum technologies can encode logical qubits within different media. We call a physical implementation of a radix-2 unit of information a physical qubit. Since today's quantum devices lack error correction, each logical qubit within an algorithm is implemented with one physical qubit in a machine. These devices, sometimes called Noisy Intermediate Scale Quantum (NISQ) devices, are error prone and are up to hundreds of qubits in size~\cite{preskill2018quantum}. Some physical qubit examples include energy levels within superconducting circuits~\cite{arute2019quantum,havlivcek2019supervised,jurcevic2021demonstration}, ions trapped by surrounding electrodes~\cite{nam2020ground,wright2019benchmarking,zhang2017observation}, neutral atoms held with optical tweezers~\cite{levine2019parallel,bernien2017probing}, and photons travelling through free space or waveguides~\cite{o2009photonic,knill2001scheme}. Each of these platforms have their unique strengths, but none has become the obvious choice for the standard quantum computing platform. 

\subsection{Physical Architectural Constraints}

Although many promising quantum technologies exist, they are imperfect and demonstrate attributes that prevent scaling in the near-term. For example, quantum state preparation, gate evolution, and measurement all are characterized by nontrivial infidelity rates and must be implemented with a limited set of instructions that depend on the control signals that are available for a specific qubit technology. Additionally, restricted \textit{connectivity} among device qubits is a key limiting factor. Qubit-qubit interaction is necessary for quantum entanglement, but unfortunately, many of today's QCs are limited in the number of qubits that are able to interact directly. Even with technologies that allow for all-to-all connectivity, such as with trapped ions, QCs are limited by the total number of qubits that are actively used during computation~\cite{pino2021demonstration}. Careful design of qubit layout on devices as well as intelligent algorithm mappers and compilers can assist with qubit communication overheads.

Qubit communication is required for quantum algorithms, but unintended communication, or \textit{crosstalk}, can cause errors during computation. Crosstalk errors experienced during quantum algorithm execution often arise from simultaneously stimulating two neighboring qubits. If the qubits are strongly coupled and ``feel'' each other, the fidelity of the two simultaneous computations can decrease. Fortunately, qubit crosstalk errors are systematic. Once identified, sources of crosstalk can be mitigated with software~\cite{murali2020software,ding2020systematic} or with improved device design~\cite{chamberland2020topological}.

Today's qubits are extremely sensitive, and as a result, crosstalk with neighboring qubits is not the only accidental signal interference that must be considered for quantum systems. Qubits also tend to couple with their surrounding environment, causing quantum information to \textit{decohere}. Amplitude damping and dephasing cause qubit decoherence errors. Amplitude damping describes the loss of energy from $\ket{1}$ to $\ket{0}$, and $T_1$ is used to denote the exponential decay time from an excited qubit state to a ground state. Dephasing error refers to the gradual loss of phase between $\ket{0}$ to $\ket{1}$ and is described by the time $T_2$. Quantum decoherence sets rigid windows for compute time on current QCs. If the qubit runtime, or period spanning the first gate up until measurement, surpasses the QC $T_1$ or $T_2$ time, the final result of computation may resemble a random distribution rather than the correct output.  

\subsection{Superconducting Circuits - A Case Study on Progress}

Superconducting circuits have emerged as a leading physical qubit implementation~\cite{majer2007coupling}. IBM has dedicated much research to the development of the Transmon-based superconducting QC, and since 2016, the company has allowed their prototype devices to be used by the public via the IBM Quantum Experience~\cite{IBMQE}. Since 2016, IBM QCs have increased from 5 to 127 qubits~\cite{heavy-hex-blog}. The significance of increasing the number of qubits on-chip to a total of $n$ is that more complex algorithms can be run that explore an exponentially larger state space, dimension $2^n$. Although the debut of larger QCs with over a hundred qubits is an impressive engineering development, limited fidelity, qubit communication, and coherence times prohibit these devices from reliably executing circuits that require all of a device's physical qubits to work in synergy.

Progress in quantum hardware is not measured simply by the consistency and count of device qubits. The gate error must also be considered as it influences the depth of circuits that a QC can run. IBM QCs have seen the average two-qubit infidelity decrease in magnitude from order $10^{-2}$ to $10^{-3}$ over the last five years~\cite{heavy-hex-blog}. Quality of multi-qubit gates is an important metric to consider when rating a QC because it enables the entanglement required for many quantum algorithms to demonstrate advantage. The importance of device size, qubit quality, and operation success is clear, thus, improving quantum systems as a whole requires a multi-targeted approach. To assist with scaling into more robust devices, IBM developed a metric referred to as Quantum Volume (QV) that aims to benchmark quantum hardware in the near term. QV is a scalar value, where higher is better, that assesses the largest random circuit of equal width and depth equal to $m$ that a QC can successfully run~\cite{cross2019validating}. As QV$=2^m$, the metric provides perspective on system performance as a general purpose QC. Capability in terms of coherence, calibration, state initialization, gate fidelity, crosstalk, and measurement quality are all captured by QV. QV is also influenced by design aspects such as chip topology, compilation tools, and basis gate set.

Recently, IBM extended the QC frontier to achieve a QV of 64, and then later 128, through the combined application of improved compiler optimizations, shorter two-qubit gates, excited state promoted readout, higher-accuracy state discrimination and open-loop correction via dynamical decoupling~\cite{jurcevic2021demonstration,IBMQS}. These breakthroughs were accompished on 27 qubit devices, demonstrating that the compute power of a quantum system is heavily influenced on compilation and low-level control; it does not depend on qubit count alone. Quantum compilation and low-level control will be discussed further in Section~\ref{software}. By focusing research to innovate in these domains, along with making developments in other areas of QC architecture, IBM anticipates to scale QCs to one million qubits and beyond~\cite{roadmap-IBM}.

\section{Programming for Pulse-level Control}
\label{software}

To target real-world quantum use cases, efficient programming languages and supporting software tool-flows are  required to represent complex classical and quantum information processing in quantum algorithms and then efficiently execute them both in simulation and on real quantum devices.
To build such an \emph{efficient} software stack, balancing between abstraction and detail is key. 
On the one hand, a transparent software stack that exposes device specifics helps programmers write tailored code, but on the other hand it dramatically increases the complexity of the toolflow. 
Considering this trade-off, advancements in building an efficient software stack have been a critical driver in pushing the field of quantum computing beyond the laboratory. 
Managing these trade-offs is particularly important for pulse-level control because it inherently requires programmer exposure to the finer details of the device. 
In this Section, we discuss state-of-the-art tools and models for quantum programming and device execution.


Quantum programming languages are designed to be user-friendly, with sophisticated control flow, debugging tools, and strong abstraction barriers between target operations and the underlying quantum hardware. Operations are thought of as ``black-box'' in the sense that the details of the physical implementation of quantum gates are hidden from the end user. This allows for modularity since a technology-independent quantum program written at the gate level can be compiled for execution on multiple QCs of different qubit types.

The most successful languages have been implemented as Python packages, such as IBM’s Qiskit~\cite{Qiskit}, Google’s Cirq~\cite{cirq_developers_2021_5182845},  Rigetti’s PyQuil~\cite{pyquil} and Xanadau's Strawberry Fields~\cite{Killoran2019strawberryfields}. Others are written as entirely new languages, such as Scaffold~\cite{scaffold} which is based on LLVM infrastructure, Quipper~\cite{quipper} which is a functional language embedded in Haskell, and Q\#~\cite{Microsoft} which is Microsoft’s quantum domain specific language.

Here, we restrict ourselves to Qiskit and provide an overview of Qiskit's programming model and its support for pulse-level control.
We encourage readers to further references on Qiskit~\cite{Qiskit} and its pulse support~\cite{pulse1,pulse2}.
Other programming frameworks and their device-level control can be reached via their corresponding webpages and documentation.

\subsection{Qiskit and the Pulse Programming Model}
Qiskit is an open-source quantum computing framework from IBM that provides tools for creating, manipulating, and running quantum programs on quantum systems independent of their underlying technology and architecture. 

The commonly used quantum programming paradigm is the circuit model. 
Such a model abstracts the physical execution of a quantum algorithm on a quantum system into a sequence of unitary gate operations on a set of qubits followed by qubit measurements. 
The gates manipulate the qubit states while measurements project these qubits onto a particular measurement basis that are extracted as classical bit-strings.
Qiskit supports this programming model via a quantum assembly language called OpenQASM~\cite{cross2021openqasm}. OpenQASM is simply an abstraction of the real quantum hardware execution in a form which is amenable to users familiar with classical programming models. 
The hardware is not capable of naively implementing or executing the quantum instructions from this model and must instead compose these operations via the control hardware.

At the device level, quantum system execution is implemented by steering the qubit(s) through a desired unitary evolution, which is achieved by careful engineering of applied classical control fields.
Thus, execution of a program represented via the quantum circuit model on a quantum system requires translating or compiling gate-level circuit instructions to a set of microwave control instructions, or pulses, which enact the desired state-transformations or measurements.
This translation is often suboptimal due to the heavy abstractions imposed across the software stack. In the circuit programming model, an atomic circuit instruction is agnostic to its pulse-level implementation on hardware, and unfortunately, the vast majority of program optimization is often done at the gate-level in the standard circuit model.
Extracting the highest performance out of quantum hardware would require the ability to craft a pulse-level instruction schedule for the optimization of circuit partitions.

Qiskit Pulse~\cite{pulse1,pulse2} was developed to describe quantum programs as a sequence of pulses, scheduled in time. 
Qiskit Pulse adds to the Qiskit compilation pipeline the capability to schedule a quantum circuit into a pulse program intermediate representation, perform analysis and optimizations, and then compile to Qiskit Pulse object code to execute on a quantum system.

To program such systems at the pulse-level in a hardware-independent manner requires the user-level instruction set to be target-compiled to the underlying system hardware components, each of which may have a unique instruction set and programming model. 
Qiskit Pulse provides a common and reusable suite of technology-independent quantum control techniques  that operate at the level of an analog stimulus, which may be remotely re-targeted to diverse quantum systems.

\begin{figure}[h!]
\fbox{
\includegraphics[width=0.95\columnwidth,trim={0cm 0cm 0cm 0cm},clip]{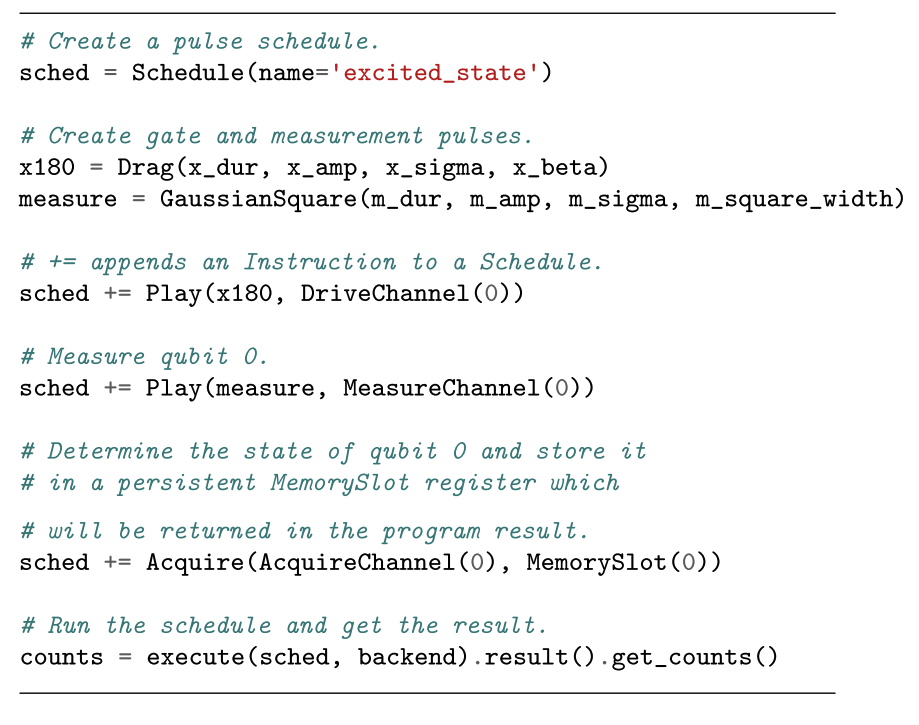}
}
\centering
\caption{A pulse schedule with deterministic instruction durations. This example prepares qubit 0 in the $\Ket{1}$ state and then measures it.~\cite{pulse2}} 
\label{QP1}
\end{figure}

Fig.~\ref{QP1} shows an example of how a pulse schedule can be explicitly coded in Qiskit. While this is a trivial example, readers can refer to ~\cite{pulse2} for an example of implementing a high-fidelity $CX$ gate based on the calibrated Cross-Resonance (CR) pulse. The CR gate is an important microwave-activated, two-qubit entangling operation that is performed by driving one qubit, the control, by the frequency of another, the target~\cite{chow2011simple}.

\begin{figure}[h!]
\fbox{
\includegraphics[width=0.95\columnwidth,trim={0cm 0cm 0cm 0cm},clip]{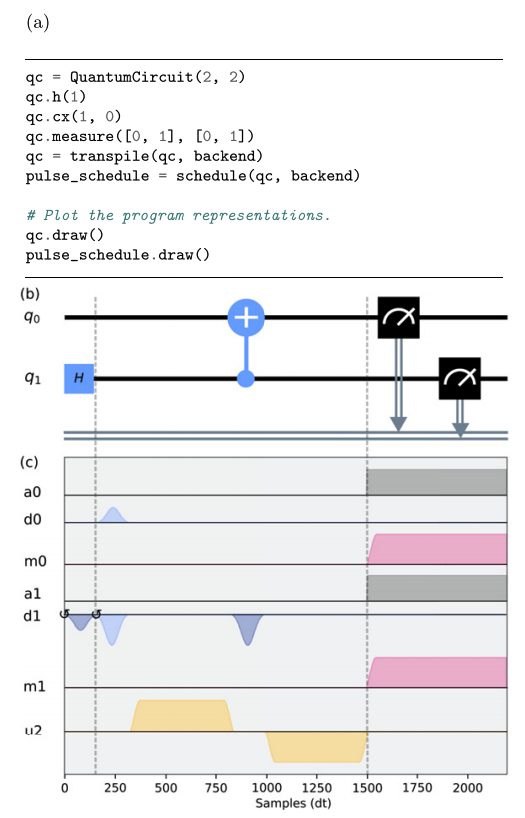}
}
\centering
\caption{(a) Qiskit code to construct a quantum circuit that prepares and measures a Bell state and then schedules the circuit to produce an equivalent pulse schedule.
(b) and (c) Visualization of the mapping between circuit instructions (b) and the composite pulse sequences that will implement the circuit elements (c).~\cite{pulse2}} 
\label{QP2}
\end{figure}

Fig.~\ref{QP2} shows provides a code snippet for scheduling a quantum circuit into a pulse schedule using Qiskit.
Qiskit Pulse users may create pulse programs to replace the default pulse programs of the native gate set provided by the backend and pass them as an argument to the scheduler.

\subsection{Pulse Support via OpenQASM3}
OpenQASM has become a de facto standard, allowing a number of independent tools to interoperate using OpenQASM as the common interchange format. 
While OpenQASM2 uses a circuit model which was described previously, quantum paradigms require going beyond the circuit model, incorporating primitives such as teleportation and the measurement model of quantum computing.
OpenQASM3~\cite{cross2021openqasm} describes a broader set of quantum circuits with concepts beyond simple qubits and gates. 
Chief among them are arbitrary classical control flow, gate modifiers (e.g. control and inverse), timing, and microcoded pulse implementations.

\begin{figure}[h!]
\fbox{
\includegraphics[width=0.95\columnwidth,trim={0cm 0cm 0cm 0cm},clip]{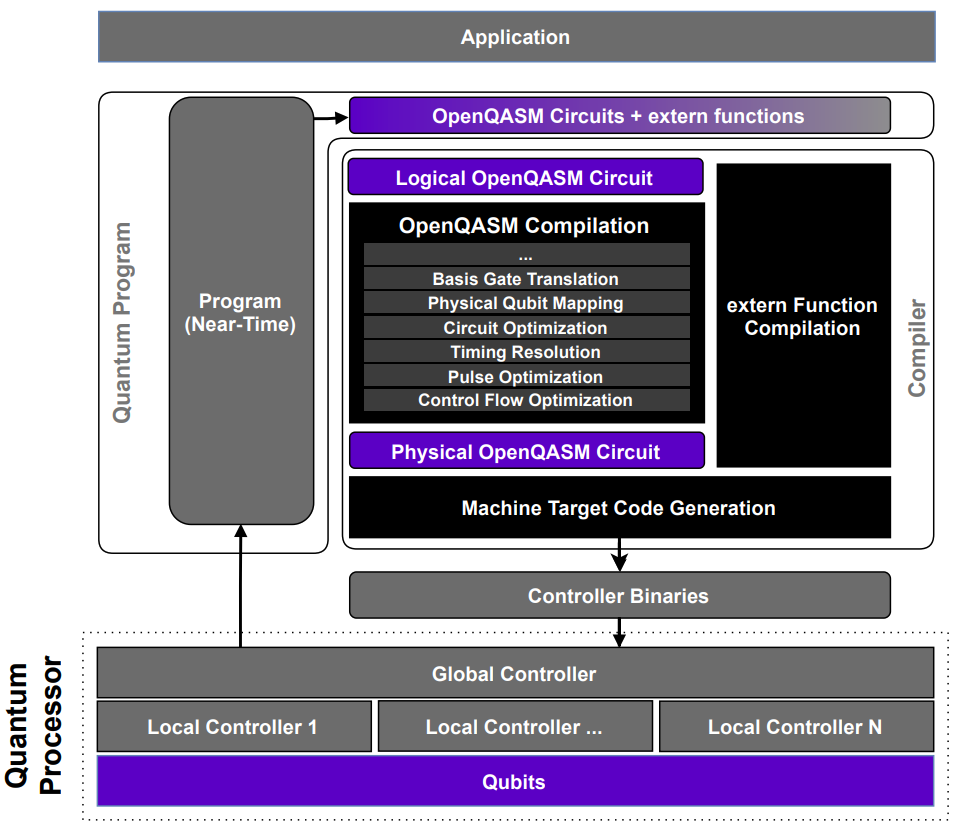}
}
\centering
\caption{The compilation and execution model of a quantum program, and OpenQASM’s
place in the flow. An OpenQASM compiler can transform and optimize all aspects of the
circuits described with the intermediate representation (IR), including basis gates used, qubit mapping, timing, pulses,
and control flow. The final physical circuit plus external functions are passed to a target code
generator which produces binaries for the quantum machine.~\cite{cross2021openqasm}} 
\label{QP3}
\end{figure}

To use the same tools for circuit development as well as for the lower-level control sequences needed for calibration, characterization, and error mitigation, it is necessary to
control timing and to connect quantum instructions with their pulse-level implementations for various qubit modalities. 
This is critical for working with techniques such as dynamic decoupling~\cite{viola1999dynamical,pokharel2018demonstration} as well as for better characterization of decoherence and crosstalk.
These are all sensitive to time and can be programmed via the timing features in OpenQASM3.
One such potential application compilation and execution flow is shown in Fig.~\ref{QP3}.

To control timing, OpenQASM3 introduces ``delay" statements and ``duration" types.
The delay statement allows the programmer to specify relative timing of operations.
Timing instructions can use the duration type which represents amounts of time measured in seconds.

OpenQASM3 includes features called ``box" and ``barrier" to constrain the reordering of gates, where the timing of those gates might otherwise be changed by the compiler. Additionally, without these directives, the desired gates could be removed entirely as a valid optimization on the logical level.

OpenQASM3 also allows specifying relative timing of operations rather than absolute timing.
This allows more flexible timing of operations, which can be helpful in a setting with a variety of calibrated gates with different durations. 
To do so, it introduces a new type ``stretch", representing a duration of time which is resolvable to a concrete duration at compile time once the exact durations of calibrated gates are known. This increases circuit portability by decoupling the circuit timing intent from the underlying pulses, which may vary from machine to machine or even from day to day.

OpenQASM3 also has added support for specifying instruction calibrations in the form of
``defcal," short for `define calibration,' declarations which allow the programmer to specify a
microcoded implementation of a gate, measure, or reset instruction.

While only a few features of OpenQASM3 are discussed here, the overall design intends to be a multilevel intermediate representation (IR), where the focus shifts from target-agnostic computation to a concrete implementation as more hardware specificity is introduced.
An OpenQASM circuit can also mix different abstraction levels by introducing constraints
where needed, but allowing the compiler to make decisions where there are no constraints.

\section{Cross-layer Compiler Optimizations for Efficient Pulse Control}
\label{Compiler}

The abstractions introduced in the layered approach of current QC stacks restrict opportunities for cross-layer optimization. 
For near-term quantum computing, maximal utilization of the limited quantum resources and reconciling
quantum algorithms with noisy devices is of importance.
Thus, a shift of the quantum computing stack towards a more vertically integrated architecture is promising.
In this Section, we discuss  optimizations that break the ISA abstraction by exposing pulse level information across the compiler stack, resulting in improvements to pulse level control as well its more efficient implementation.

\begin{figure}[h!]
\fbox{
\includegraphics[width=0.95\columnwidth,trim={0cm 0cm 0cm 0cm},clip]{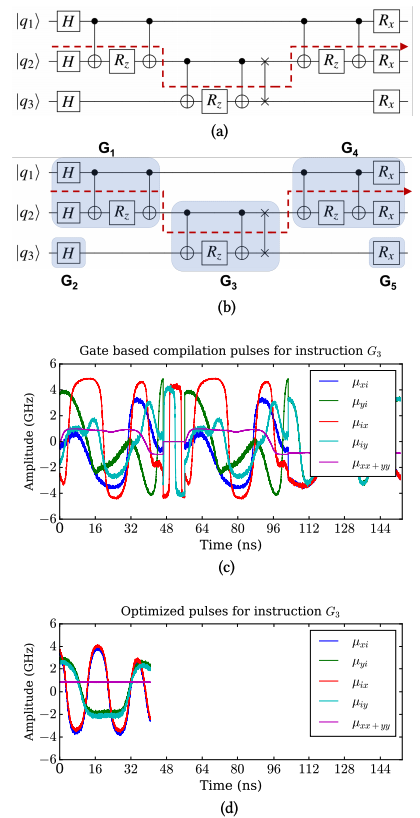}
}
\centering
\caption{Example of a QAOA circuit demonstrating the difference between gate-based compilation and the novel compilation methodology. (a) Standard circuit (red arrow indicates the critical path). (b) Circuit with aggregated instructions. (c) Standard compilation pulses for G3. (d) Aggregated compilation pulses for G3. Each line represents the intensity of a control field. The pulse sequence in (d) is much shorter in duration and easier to implement than that of (c).~\cite{Shi2019}} 
\label{QP4}
\end{figure}

\subsection{Optimized Compilation of Aggregated Instructions for Realistic Quantum Computers}
The work~\cite{Shi2019} proposes a quantum compilation technique that optimizes for pulse control by breaking across existing abstraction barriers.
Doing so reduces the execution latency while also making optimal pulse-level control practical for larger numbers of qubits.
Rather than directly translating one- and two-qubit gates to control pulses, the proposed framework first aggregates these small gates into larger operations.
Then the framework manipulates these aggregates in two ways. First, it finds commutative operations that allow for much more efficient schedules of control pulses. Second, it uses quantum optimal control on the aggregates to produce a set of control pulses optimized for the the underlying physical architecture. 
In all, the technique greatly exploits pulse-level control, which improves quantum efficiency over traditional gate-based methods.
At the same time, it mitigates the scalability problem of quantum optimal control methods. 
Since the technique is software-based, these results can see practical implementation much faster than experimental approaches for improving physical device latency. 
Compared to traditional gate-based methods, the technique achieves execution mean speedup of 5x with a maximum speedup of 10x.

Two novel techniques are implemented:
a) detecting diagonal unitaries and scheduling commutative instructions to reduce the critical path of computation; 
b) blocking quantum circuits in a way that scales the optimal control beyond 10 qubits without compromising parallelism.

For quantum computers, achieving these speedups, and thereby reducing latency, is do-or-die: if circuits take too long, the qubits decohere by the end of the computation. 
By reducing the latency 2-10x, this work provides an accelerated pathway to running useful quantum algorithms, without needing to wait years for hardware with 2-10x longer qubit lifetimes.
An illustrative example is shown in Fig.~\ref{QP4}.

\begin{figure}[h!]
\fbox{
\includegraphics[width=0.95\columnwidth,trim={0cm 0cm 0cm 0cm},clip]{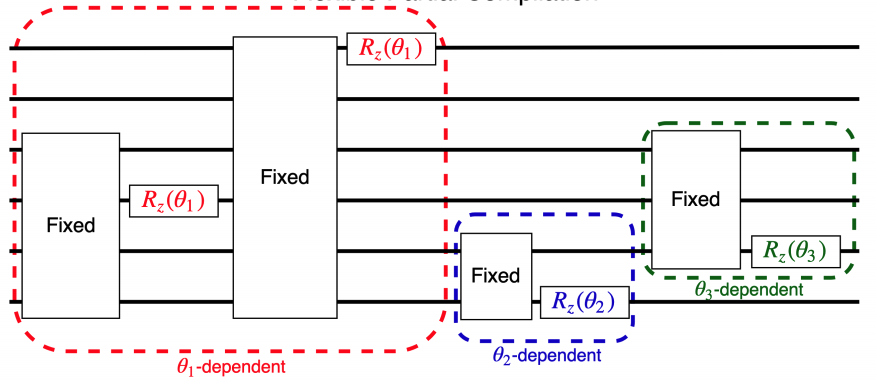}
}
\centering
\caption{Flexible partial compilation blocks the circuit into subcircuits that depend on exactly one parameter. Hyperparameter optimization is used to precompute good hyperparameters (learning rate and decay rate) for each subcircuit. When gate angles are specified at runtime, the tuned hyperparameters quickly find optimized pulses for each subcircuit.~\cite{Gokhale2019}} 
\label{QP5}
\end{figure}

\subsection{Partial Compilation of Variational Algorithms for Noisy Intermediate-Scale Quantum Machines}

Each iteration of a variational algorithm depends on the results of the previous iteration–hence, the compilation must be interleaved through the computation. 
The noise levels in current quantum machines and the complexity of the variational use cases that are useful to solve, result in a very complex parameter tuning space for most algorithms.
Thus, even small instances require thousands of iterations.
Considering that the circuit compilation is on the execution critical path and cannot be hidden, the compilation latency for each iteration becomes a serious limitation. 

To cope with this limitation on compilation latency, past work on VQAs has performed compilation under the standard gate-based model. 
This methodology has the advantage of extremely fast compilation – a lookup table maps each gate to a sequence of machine-level control pulses so that the compilation simply amounts to concatenating the pulses corresponding to each 
gate. 

The gate-based compilation model is known to fall short of the GRadient Ascent Pulse Engineering (GRAPE)~\cite{GRAPE1,GRAPE2} compilation technique, which compiles directly to the level of the machine-level control pulses that a QC actually executes. 
As has been the theme of this paper, the pulse-level control provides for considerably more efficient execution on the quantum machine. 
GRAPE has been used to achieve 2-5x pulse speedups over gate-based compilation for a range of quantum algorithms, resulting in lower decoherence and thus increased fidelity.

However, GRAPE-based compilation has a substantial cost: compilation time. 
This would potentially amount to several weeks or months of total compilation latency during thousands of iterations since millions of iterations will be needed for larger problems of significance. By contrast, typical pulse times for quantum circuits are on the order of microseconds, so the compilation latency imposed by GRAPE is untenable. 

This proposal~\cite{Gokhale2019} introduces the idea of partial compilation, a strategy that approaches the pulse duration speedup of GRAPE, but with a manageable overhead in compilation latency. 
This powerful new compiler capability enables a realistic architectural choice of pulse-level control for more complex near-term applications.
Two variations are proposed: a) strict partial compilation, a strategy that pre-computes optimal pulses for parametrization-independent blocks of gates; and b) flexible partial compilation, a strategy that performs as well as full GRAPE, but with a dramatic speedup in compilation latency via precomputed hyperparameter optimization.
An illustration of the flexible partial compilation is shown in Fig.~\ref{QP5}.

\begin{figure}[h!]
\fbox{
\includegraphics[width=0.95\columnwidth,trim={0cm 0cm 0cm 0cm},clip]{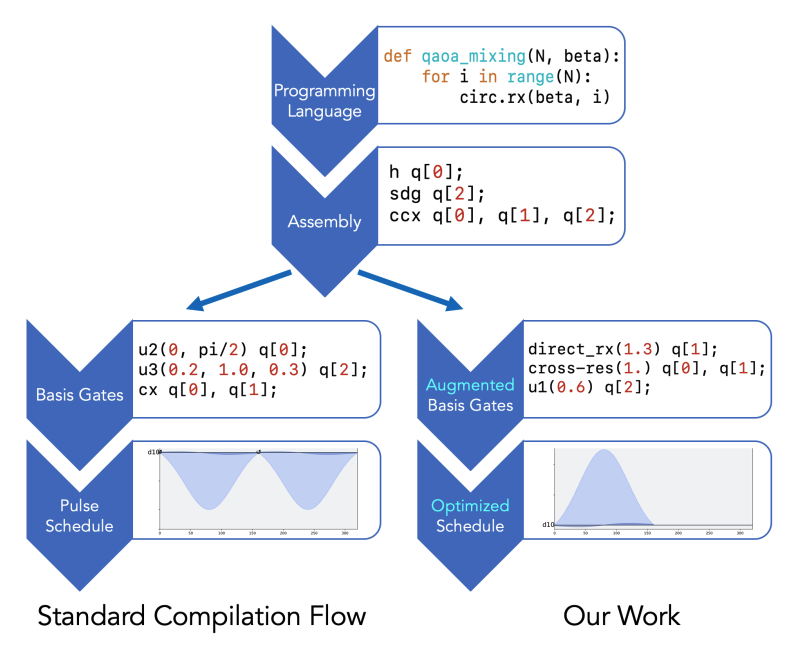}
}
\centering
\caption{Like classical programs, quantum programs undergo a compilation process from high-level programming
language to assembly. However, unlike the classical setting,
quantum hardware is controlled via analog pulses. This work optimizes the underlying pulse schedule by augmenting the set basis gates to match hardware. The compiler automatically optimizes user code, which therefore remains hardware-agnostic.~\cite{gokhale2020}} 
\label{QP6}
\end{figure}

\subsection{Optimized Quantum Compilation for Near-Term Algorithms with Qiskit Pulse}

While pulse optimization has shown promise in previous quantum optimal control (QOC), noisy experimental systems are not entirely ready for compilation via QOC approaches. 
This is because QOC requires an extremely accurate model of the machine, i.e., its Hamiltonian.
Hamiltonians are difficult to measure experimentally and moreover, they drift significantly between daily recalibrations. 

Experimental QOC papers incur significant pre-execution calibration overhead to address this issue. 
By contrast, this work~\cite{gokhale2020}, \cite{gokhale2021faster} proposes a technique that is bootstrapped purely from daily calibrations that are already performed for the
standard set of basis gates. 
The resulting pulses are used to create an augmented basis gate set. 
These pulses are extremely simple, which reduces the control error and also preserves intuition about underlying operations, unlike traditional QOC. 
This technique leads to optimized programs, with mean 1.6x error reduction and 2x speedup for near-term algorithms.
The proposed approach can target any underlying quantum hardware.
An overview is shown in Fig.\ref{QP6}

Four key optimizations are proposed, all of which are enabled by pulse-level control:
(a) access to pulse-level control allows implementing any single-qubit operation directly with high fidelity, circumventing inefficiencies from standard compilation;
(b) although gates have the illusion of atomicity, the true atomic units are pulses. The proposed compiler creates new cancellation optimizations that are otherwise invisible;
(c) Two-qubit operations are compiled directly down to the two-qubit interactions that the hardware actually implements;
(d) Pulse control enables $d$-level qudit operations, beyond the 2-level qubit subspace.

\section{Simulation of Pulses and with Pulses}

In this Section, we highlight some works on simulation.
There are two forms of simulation which are relevant.
The first is classical simulation of quantum devices to better understand device behavior.
The second is simulating the quantum physical aspects of a complex system on a quantum machine itself.
Both of these simulation domains intersect with pulse-level control.
On the one hand, the effective classical simulation of quantum devices will require accurate capture of pulse level device phenomena.
On the other hand, effectively modeling complex quantum physical systems also requires precise control of execution on the quantum device, which is enabled by pulse-level control.

\subsection{Classical Simulation: Capturing the Time-varying Nature of Open Quantum Systems}
Classical simulation of quantum devices is critical to better understand device behavior. This is especially true in the near-term for noisy prototype architectures. Simulation tools accomplish a variety of different tasks including modeling noise sources, validating calculations and designs, and evaluating quantum algorithms. Since quantum simulation tools are classical, they often require supercomputers to simulate quantum devices that are large in terms of either qubit count or hardware components they contain. Quantum simulators imitate the operation of quantum devices and can be used to build the Hamiltonians required to derive control pulses for the qubit or qudit space. As a result, quantum simulators are an important tool in control engineering because complex quantum operations require the generation of fine-tuned drive signals.

Quantum systems are dynamic and evolve with time. As a result, simulation tools for quantum mechanical systems need to take variation into consideration, whether that variation stems from intentional gate application or from unintentional drift or environmental coupling. QuaC, or ``Quantum in C,'' was developed to capture the time-varying nature of open quantum systems, including realistic amplitude damping and dephasing along with other correlated noises and thermal effects~\cite{otten2017quac}. The simulator is not limited to a specific type of qubit or qudit technology, and although QuaC is classical simulator, the tool can model quantum systems with enough granularity to develop pulses required for low-level control. QuaC has been used for many applications, such as in the discovery of improved error models to better understand noisy quantum systems~\cite{otten2019recovering} and for the study of the coupling required between quantum dots and photonic cavities for entanglement transfer~\cite{otten2019optical}. Additionally, QuaC was applied in the comparison of different quantum memory architectures to discover optimum features, such as encoding dimension~\cite{otten2021impacts}.   

\subsection{Quantum Simulation: Hardware Efficient Simulation with Qiskit Pulse}
The proposal in~\cite{stenger2021} simulates a quantum topological condensed matter system on an IBM quantum processor.
The simulation is done within the qubits’ coherence times using pulse-level instructions provided by Qiskit Pulse. 
Ideally, capturing the system characteristics would require simulation by continuous time-evolution of qubits under the appropriate spin Hamiltonian obtained from a transformation of the fermion Hamiltonian. 
However, practically, this is run on quantum devices by having this ``analog" simulation decomposed and mapped onto the calibrated native basis gates of a quantum computer, making it ``digital". 
This digital implementation on noisy quantum hardware limits precision and flexibility and is thus unable to avoid the accumulation of unnecessary errors. 
Pulse-level control improves this by allowing for a ``semi-analog" approach.
The proposal shows a pulse-scaling technique that, without additional calibration, gets closer to the ideal analog simulation.

Topologically-protected quantum computation works by moving non-Abelian anyons, such as Majorana zero modes (MZMs), around each other in two dimensions to form three-dimensional braids in space-time~\cite{majorana}.
Thus far, there has been no definitive experimental evidence of braiding due to dynamical state evolution~\cite{majorana2}

This work simulates a key part of a topological quantum computer: the dynamics of braiding of a pair of MZMs on a trijunction. 
Braiding is implemented by parametrically adjusting the Hamiltonian parameters; the time evolution is implemented using the Suzuki-Trotter decomposition, with each time step implemented by one- and two-qubit gates. 
Fidelity is significantly improved by using pulse-level control to scale cross resonance (CR) gates derived from those pre-calibrated on the backend, thereby enabling coupling of qubit pairs with shorter CR gate times.
Specifically, using native $CX$ gates, only 1/6th of a full braid can be performed.
Whereas, using the pulse-enabled scaled gates leads to performing a complete braid.

\section{Engineering Optimal Pulses}
\label{Control Engineering}

\subsection{Error-robust Single-qubit Gate Set Design}

The work~\cite{Carvalho2021} proposes analog-layer programming on superconducting quantum hardware to implement and test a new error-robust single-qubit gate set.
To build this, analog pulse waveforms are numerically optimized via the use of Boulder Opal, a custom Tensorflow-based package~\cite{QCtrl}.
The optimized pulses enact gates which are more resilient against dephasing, control-amplitude fluctuations, and crosstalk. Before the implementation in real hardware, these pulses go through a calibration protocol that can be fully automated~\cite{QCtrl_calibration} to account for small distortions that may happen in the control channels. 
The experiments are performed on IBM quantum machines and programmed via the Qiskit Pulse API, which translates the pulses designed using Boulder Opal into hardware instructions. 

The experiments show that when pulses are optimized to be robust against amplitude or dephasing errors, they can outperform the default calibrated Derivative Removal by Adiabatic Gate (DRAG) operations under native noise conditions. 
These optimized pulses are built including both a 30-MHz-bandwidth sinc-smoothing-function and temporal discretization to match hardware programming.

Using optimized pulses, single-qubit coherent error rates originating from sources such as gate miscalibration or drift are reduced by up to an order of magnitude, with average device-wide performance improvements of 5x. 
The same optimized pulses reduce the gate error variability across qubits and over time by an order of magnitude.

\subsection{Reinforcement Learning for Error-Robust Gate Set Design}

As discussed above, it has been demonstrated that the use of robust and optimal control techniques for gateset design can lead to dramatic improvements in hardware performance and computational capabilities. 

The design process is straightforward when Hamiltonian representations of the underlying system are precisely known, but is considerably difficult in state-of-the-art large-scale experimental systems. 
A combination of effects introduces challenges not faced in simpler systems including unknown and transient Hamiltonian terms, control signal distortion, crosstalk, and temporally varying environmental noise. 
In all cases, complete characterization of Hamiltonian terms, their dependencies, and dynamics becomes unwieldy as the system size grows.

\begin{figure}[h!]
\fbox{
\includegraphics[width=0.9\columnwidth,trim={0cm 0cm 0cm 0cm},clip]{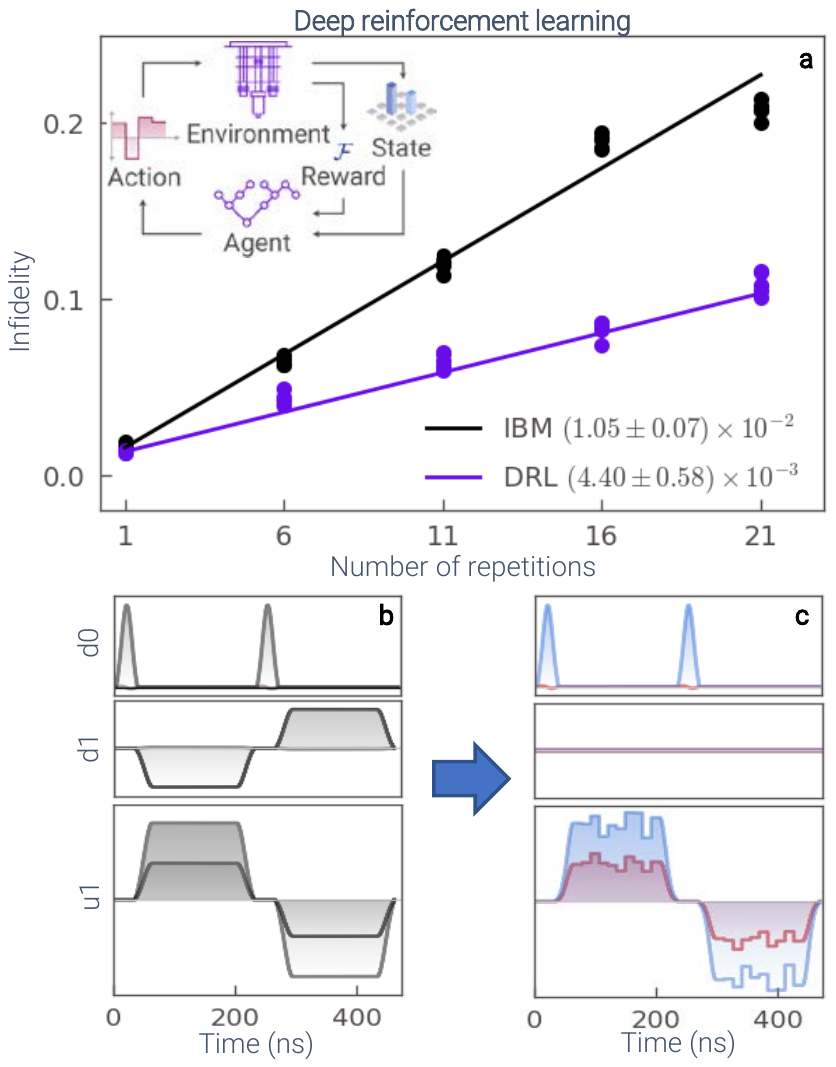}
}
\centering
\caption{Optimization of a $ZX(-\pi/2)$ entangling gate for a superconducting device~\cite{Baum2021} using a deep reinforcement learning (DRL) optimization routine (see inset for a diagram of the optimization cycle).  
(a) Infidelity measurement after different numbers of repeated application of the gate for the IBM default and DRL optimized gate. Approximate gate error is extracted from the slope of the infidelity data. (b-c) Waveforms for the (b) IBM default and (c) DRL optimized $ZX(-\pi/2)$ gates. Note that channel d1, used as a cancellation tone in the IBM default is not used in the optimized gate. 
} 
\label{qctrl_RL}
\end{figure}

The work~\cite{Baum2021} proposes a black-box approach to designing an error-robust universal quantum gate set using a deep reinforcement learning (DRL) model, as shown in the inset of Fig.~\ref{qctrl_RL}(a).
The DRL agent is tasked to learn how to execute high fidelity constituent operations which can be used to construct a universal gate set.
It iteratively constructs a model of the relevant effects of a set of available controls on quantum computer hardware, incorporating both targeted responses and undesired effects. 
It constructs an $RX$($\pi/2$) single-qubit driven rotation and a $ZX$($-\pi/2$) multi-qubit entangling operation by exploring a space of piecewise constant (PWC) operations executed on a superconducting quantum computer programmed using Qiskit Pulse.

The constructed single-qubit gates outperform the default DRAG gates in randomized benchmarking with up to a 3x reduction in gate duration. 
Furthermore, the use of DRL defined entangling gates within quantum circuits for the SWAP operation  shows 1.45x lower error than calibrated hardware defaults. 
These gates are shown to exhibit robustness against common system drifts, providing weeks of performance without needing intermediate recalibration.

\section{Advanced Architectures with Cavity Systems}

Improved characterization and understanding of quantum devices leads to improved low-level control, and this opens the door to discovering novel hardware use that pushes the state-of-the-art in quantum computing forward. As the level of precision with which we manipulate quantum devices increases, fine-tuned drive signals become available for the realization of complex Hamiltonians and gates. 

This Section includes a discussion of how the codesign of quantum hardware and low-level control has enabled innovative architectures based on cavities. Cavity-Transmon architectures are particularly exciting for quantum computing as they have been proposed for use in quantum memory and error correction schemes. 

\subsection{Robust Quantum Control over Cavity-Transmon Systems}

Oscillator cavities with photonic or phononic modes have long coherence times extending to tens of milliseconds, but unfortunately, these devices are difficult to control, prohibiting the generation of arbitrary quantum states needed for quantum computation~\cite{heeres2015cavity}. However, if the storage cavity is coupled directly to a Transmon qubit, quantum states can be prepared in the Transmon and swapped into the storage cavity for universal control. After the quantum SWAP occurs, the oscillator holds a bosonic qubit. This idea led to the development of the superconducting cavity qubit module. 

Superconducting cavity qubit modules are built using superconducting circuits comprised of a Transmon qubit, a storage cavity, and a readout cavity connected to a coupler port \cite{paik2011observation,arrangoiz2019resolving,sletten2019resolving}. In this system, the Transmon qubit is used for quantum information processing, while the storage cavity, or the oscillator, can interact with the Transmon to encode state within an oscillator mode~\cite{krastanov2015universal}.  These systems are characterized by a long-lived superconducting storage cavity coherence, a large Hilbert space for representing states, and fast, high-fidelity 
measurement and readout of the qubit state. All of these features make cavity qubit modules an attractive choice for storing encoded qubits. They also have potential within future quantum error correction protocols.

The work in~\cite{krastanov2015universal,heeres2015cavity} presents the SNAP gate that improves previous efforts of Transmon-cavity interaction for quantum computation. This gate efficiently enables the construction of arbitrary unitary operations, offering a scalable path towards performing quantum computation on qubits encoded in oscillators. Combined cavity and Transmon drive pulses provide control of the system to implement the SNAP gate. These pulses can be generated by gradient-based optimum control, or GRAPE~\cite{ma2021quantum}.

The superconducting cavity module consisting of a storage cavity and a Transmon qubit is an example of an ancilla-assisted system. The main goal of the system to take advantage of the long coherence times of oscillator modes. The challenge of limited control associated with the storage cavity is alleviated through Transmon qubit coupling. Employing a Transmon in the cavity qubit module, however, does not come without cost. The ancilla qubit injects noise to the system due to shorter coherence windows associated with the Transmon (tens of microseconds). Thus, there is room to further improve the quantum control of the superconducting cavity module. Quantum error correction techniques are required to protect cavity systems from ancilla-introduced errors. Fortunately, a solution based on path independence has been proposed to develop fault tolerant quantum gates that are robust to ancilla errors~\cite{ma2020path,reinhold2020error}.

\subsection{Specialized Architectures for Error Correction}

Quantum control allows the desired quantum device behavior from carefully designed classical signals. Control techniques that continue to reduce error in near term QCs are still under development, but it must be kept in mind that the goal of quantum science is to eventually build architectures that can implement fault tolerance. As a result, new device architectures and supporting control systems should be designed concurrently. 

Looking forward, there are many design bottlenecks that must be addressed to scale the current qubit technology. For example, superconducting qubits face issues associated with crosstalk between qubits, limited area for control wires, and inconsistencies during fabrication. These challenges must be sidestepped, and one approach is to reduce the amount of hardware scaling required to minimize the burden on engineering and material science efforts. 

New architectures based on superconducting cavity technology have been proposed that aim to implement fault tolerance while reducing the requirements of the physical hardware~\cite{duckering2020virtualized}. With reduced hardware, some of the challenges associated with scaling quantum control mechanisms will be resolved. 

The approach in~\cite{duckering2020virtualized} aims to move towards scalable fault tolerant architectures by combining compute qubits with memory qubits for a 2.5D device design. The combined compute and memory unit will be constructed with a Transmon coupled with a superconducting cavity. The cavity is characterized by coherence times that are much longer than those of the Transmon qubit. This advantage allows the cavity memory unit coupled to the Transmon compute unit to be used for random access to error-corrected, logical qubits stored across different memories, creating a simple virtual and physical address scheme. Error correction is performed continuously by loading each qubit from memory. The error correction used in~\cite{duckering2020virtualized} includes two efficient adaptations of the surface code, Natural and Compact. 

The 2.5D architecture has many architectural advantages that provide avenues for simplified control engineering. First, the combined Transmon and cavity module requires only one set of control wires. Compared to traditional 2D architectures that require dedicated control wires and signal generators, there is potential to reduce the amount of control hardware by a factor of $n$ where $n$ is the number of modes, and thus the number of qubits, that can be stored in the storage cavity. Additionally, the 2.5D architecture allows for transversal $CX$ operations that can extend into the z-plane of the cavity to act on qubits stored within modes. This $CX$ operation allows for lattice surgery operations with improved connectivity between logical qubits and that can be executed 6x faster than standard lattice surgery $CX$ operations. Faster gates are a huge win for device control when qubits are constrained by a coherence time.

\section{Applying Qudits for Acceleration}

Improved low-level control has increased access to higher-dimensional encoding. In this Section, we describe work that employs qudits for quantum processing gains during gate operation and entangled state preparation. 

\subsection{Toffoli Gate Depth Reduction in Fixed Frequency Transmon Qubits}

Quantum information encoding is typically binary, but unlike classical computers, most QCs have multiple accessible energy levels beyond the lowest two used to realize a qubit. Accidental use of these upper energy levels can insert errors into computation. However, these energy levels that transform a qubit into a qudit can be intentionally used with careful quantum control to achieve performance gains during computation. Control pulses that access the qudit space must be carefully designed, taking hardware constraints such as shorter decay time associated with higher energy levels into consideration~\cite{blok2021quantum}.

Implementing high-dimensional encoding for quantum information provides the benefit of data compression since qubit states can be represented with qudit formalism. Condensed information storage with qudits has been proposed for use in efficient applications of quantum error correction~\cite{muralidharan2017overcoming}, communication protocols~\cite{vaziri2002experimental}, and cryptography~\cite{bruss2002optimal}. Another exciting application for high-dimensional encoding is in gate depth reduction for multi-qubit gates~\cite{fedorov2012implementation,gokhale2019asymptotic}. Multi-qubit gates are required in many quantum algorithms, but they must be decomposed into smaller operations that agree with both the basis gate library and connectivity graph of a QC. These decompositions can be costly in terms of total single- and two-qubit operations, so opportunities for gate depth reduction can directly improve the overall circuit fidelity on today's noisy QCs. 

Qiskit Pulse allows users to access of higher energy levels on their Transmon-based QCs~\cite{IBM-high-dim}. Using this low-level control of the hardware,  gates for qutrits, or radix-3 quantum information that uses the basis states of $\ket{0}$, $\ket{1}$, and $\ket{2}$, can be defined. Recent work using Qiskit Pulse experimentally demonstrated that extending into the qutrit space can provide advantages for qubit-based computing~\cite{galda2021toffoli,galda2021implementing}. In these schemes, qudits were used for intermediate computation in gate realization. The overall algorithm, however, still processes in radix-2, maintaining its ``qubits in, qubits out'' structure. 

In~\cite{galda2021implementing}, the authors propose a Toffoli gate decomposition that is improved by intermediate qutrits. The Toffoli gate, or controlled-$CX$ ($CCX$), is an important quantum gate that has been targeted by optimal control techniques since it is widely used in reversible computation, error correction, and chemistry simulation among other quantum applications~\cite{kim2021high}. The authors introduce a decomposition that uses single- and two-qutrit gates to achieve an order-preserving, Toffoli decomposition that only requires four, two-Transmon interactions. The value of being order preserving is that the operation is friendly to near-term devices that demonstrate nearest-neighbor connections. 

Ref.~\cite{galda2021implementing} describes the control pulses required to realize the qutrit operations on IBM hardware. The successful demonstration of this Toffoli decomposition provides a significant gate reduction from the optimum qubit-based alternative that requires eight, two-Transmon interactions. Average gate fidelity of qutrit Toffoli execution was measured to be about 78\%. The qutrit Toffoli gate was benchmarked against the optimum qubit-based decomposition showing a mean fidelity improvement of around 3.82\% and an execution time reduction of 1$\mu s$. Additional error correction techniques via quantum control measures, such as dynamical decoupling, were also employed for further performance gains with the qutrit Toffoli.

\subsection{High-dimensional GHZ Demonstration}

Quantum entanglement is a hallmark that distinguishes quantum computation from classical processing techniques. It has a wide range of applications from secure communication protocols to high-precision metrology. There have been many experimental demonstrations of entanglement, but they have primarily focused on radix-2 systems. An advantage of exploring higher-dimensional entanglement includes increased bandwith in quantum communication protocols, such as those for superdense codes and teleportation.

At least two systems are needed for quantum entanglement, and a GHZ state is a special type of entanglement, called multipartite entanglement, that is shared between three or more subsystems, i.e. qubits or qudits. High-dimensional entanglement has been demonstrated on numerous occasions with photonic devices~\cite{erhard2020advances}, but with careful pulse-level control using Qiskit Pulse, a qutrit GHZ state shared between three qudits was prepared on IBM Transmon QC~\cite{cervera2021experimental}. This demonstration was the first of its kind on superconducting quantum technology.

In~\cite{cervera2021experimental}, the GHZ circuit for three qutrits was built using IBM calibrated gates, $RY(\theta)$ and $CX$, along with specially programmed and calibrated single-qutrit gates. To detect qutrit states, a custom discriminator was developed that is sensitive to three basis states for during measurement. Experiments to generate entanglement were ran using IBM Quantum Cloud resources, and GHZ states were produced around 30000 times faster than leading photonic experiments~\cite{erhard2018experimental}. Tomography confirmed the fidelity of entangled state preparation on a five-qubit IBM machine to be approximately 76\%. Additionally, the three-qutrit GHZ state was verified further using the entanglement witness protocol.

\section{Expanding the Quantum Hardware Community}
\label{expanding-q-hardware-community}

Exciting breakthroughs have been seen in the NISQ era, but reaching the full promise of quantum-accelerated chemical simulation, data processing, and factoring, will require large-scale and preferably fault tolerant quantum computers~\cite{preskill2018quantum}. Refinement of algorithms, devices, and processes is still necessary to unlock the computational advantages of quantum information processing. Unfortunately, it is anticipated that the demand for a quantum workforce to pursue these goals will greatly outweigh the supply in the near term as the amount of individuals pursuing advanced degrees in the traditional background of quantum physics and engineering is not increasing~\cite{fox2020preparing}. Thus, it is of critical importance to 1) develop a quantum-aware workforce starting at all age ranges and educational backgrounds in STEM and 2) improve the general population's overall understanding of quantum technology so that QC frontiers continue to advance.
As end-to-end quantum solutions mature, it is imperative to assemble a multidisciplinary network with a broad set of skills that is ready to face all challenges relating to quantum scalability. 


\subsection{Quantum Hardware Education}

Reaching fault tolerant QCs is heavily dependent on dedicating adequate resources to develop improved hardware and control. This involves building a quantum community with skill sets that range from quantum aware to quantum specialist. A first step to expanding the quantum hardware community is limiting the barriers to entry. The quantum net could be cast to a wider audience by increasing efforts to introduce concepts related to quantum computing and technology at all education levels. Additionally, expanding quantum optimal control research will include engaging a diverse set of willing students of all ages. STEM has largely overlooked many demographic groups, but as quantum research is still in its infancy compared to many classical fields, there is opportunity to ensure that engineers and scientists from underrepresented groups feel a sense of belonging in the quantum community and feel that they can contribute to its growth. Expanding diversity and inclusion in quantum computing will continue to advance the field in both the academic and industrial setting.

In this Section, we will discuss three key strategies to allow quantum information to reach a wider audience by tailoring instruction to different levels of education. First, if students become more familiar with core quantum concepts at an early age, they are better prepared to pursue careers in quantum hardware and quantum control. Second, as the demand for a quantum workforce increases, higher education will gain a better perspective of what is required for degree plans in quantum science at both the undergraduate and graduate levels. Finally, it is possible to train seasoned scientists and engineers from classical areas of study so that their refined expertise can be applied to the research and development of quantum technology.

\subsubsection{Early Education, K-12}

Introducing quantum information concepts at early ages might assist with lowering the barrier to enter a career in quantum science. The theory behind quantum computing contains significant mathematics content, but it is possible to introduce core concepts in a non-mathematical manner. For example, educators at this level could work towards demystifying the field, focusing on the ``what'' rather than emphasizing and theoretically proving the reasoning behind purely quantum phenomena. It is often through the attempt to deliver a crash-course of the complex answers for ``how'' quantum works that leaves many students lost. For example, rather than teaching the details of superposition and its implications on quantum states, young learners can be taught at a higher level to get comfortable with the idea that some items, like the the combination of spoon and fork to make the spork, can be two things at once~\cite{K-12-quantum}.

Recent advances in quantum science are largely caused by improvements made to supporting technology. For example, without the parallel development of cryogenic technology, many quantum computing platforms would not support computation due to the lack of a stable operational environment. Similar strides in refining quantum control have also allowed QCs to become more robust.
Indirectly, concepts related to the fundamental nature of quantum hardware and optimal control to the development of quantum informatics can be introduced to a younger audience. Scientific discovery includes many moving parts that at first glance, may not seem significant, but are a critical part of development and experimentation. Everyday, tools and methods are used to accomplish a bigger goal, just as quantum hardware and optimum control enable quantum computing, and younger scientists can understand this by thinking about all of the equipment and instructions needed to complete a simple task or experiment. As a basic example, a cake cannot result from ingredients without the use of bowls, measuring cups, spoons, a working oven, and a detailed recipe for the baker to apply.

\subsubsection{Higher Education}

Select undergraduate colleges and universities along with graduate schools offer exposure to coursework in quantum computing and quantum technologies. These programs, however, expand to clearly define the quantum workforce pipeline so that its output can be maximized to meet the needs of academia, government, and industry.

When considering education in the space of quantum informatics, a doctorate in quantum physics immediately comes to mind. Such expert individuals, however, are in limited supply while there is projected to be a substantial demand for a quantum-ready workforce. Additionally, not every quantum scientist needs to be well-versed in algorithms and theory. The future workforce is expected to be populated by individuals with varied backgrounds and levels of training, as opportunities for employment within the quantum industry can range drastically. More technical roles range from engineer (i.e., electrical, software, optical, systems, materials, etc.), experimental scientist, theorist, technician, and application researcher. Less technical roles include those handling business development, sales, and legal strategies related to emerging technologies. Because of the wide breadth of backgrounds valuable to quantum computing, all of which necessary to further progress on quantum devices and control, many have considered how to structure quantum-related majors and minors at the undergraduate level. These types of programs would help satisfy the demand of quantum-aware and quantum-proficient professionals in the workforce. 

Focusing on the more technical portion of the workforce, quantum training programs could easily take advantage of existing programs relating to engineering, computation and science. Training in dynamics, chemistry, electromagnetism, and programming languages are just a few examples of courses that would supply individuals with the skills needed to contribute to the field of quantum. Along with a general education, application-specific areas of quantum, such as devices, algorithms, software, and protocols, would be taught depending on the student's targeted career path. In depth discussions related to the development of quantum engineering programs can be found in~\cite{dzurak2021development,asfaw2021building,fox2020preparing}.

\subsubsection{Continuing Education as a Professional}

In quantum education, much emphasis has been placed on algorithms, but this does not reflect on the breadth of needs for quantum research in industry and academics. In actuality, a much more multifaceted group of researchers composed of different backgrounds is needed to push quantum technology toward the ultimate goal of fault tolerance.  For example, many strides in quantum technology can be made by recruiting from the classical fields of science as many classical problems appear on the roadmap to scalable quantum technology. There are many ways in which quantum devices can improve, but some key areas include the development of better fabrication techniques, more efficient refrigeration technologies, and more accurate control methods and hardware. These example design problems could be approached by engineers that are classically trained but are literate in the basics of quantum science.


Industry constantly evolves as new technology appears. Thus, it is important that experienced professionals continue to explore their horizons to develop their skills and keep their knowledge up to date. In the space of quantum informatics, there are many resources to introduce seasoned scientists and engineers to core quantum concepts in a way that is accessible. There are many tools, such as the Qiskit Textbook~\cite{Qiskit-textbook}, blogs~\cite{q-aviary}, workshops~\cite{I2Q}, and online courses~\cite{edx-course}, that can allow individuals to explore quantum computing at their own pace. When experienced professionals dive into quantum computing, a greater pool of knowledge to assist with developing new and scaling current quantum methods and hardware results.

\subsection{Improving Quantum Technology Awareness}

Many exciting developments in the field of quantum information processing are unfortunately buried in research papers, specialized conference sessions, and meetings that are inaccessible to those not immediately working on the issues at hand. So while quantum computing holds great potential, there is the risk that the emerging technology could be undervalued or worse, misunderstood. A goal of expanding the quantum hardware community is to increase the public's awareness about the reasonable expectations and timelines to have for QCs. By improving communication channels and trust between the QIS community and the rest of the world, more individuals may consider how quantum research may impact their lives or even how they can get involved. Initial steps such as research groups producing news articles describing theoretical and experimental breakthroughs in layperson's terms could make a huge difference so that quantum progress reaches a wider audience is understood with greater accuracy.

\section{Conclusion}
\label{Conclusion}


Quantum systems evolve in a continuous manner, and their underlying low-level control signals are continuous as well.
Thus, the pulse-based quantum computing approach utilizing continuous control signals potentially offers a much richer and more flexible use than the highly popularized gate-based approach.
The ability to engineer a real-time system Hamiltonian allows us to navigate the quantum system to the quantum state of interest through generating accurate control signals. While the benefits to the pulse approach are clear, there are several challenges stemming from the inherent complexities of a full-stack approach, including, but not limited to, defining the machine Hamiltonian, outlining the programming model, overheads from optimization and compilation, and both classical and quantum simulation capabilities. 
Overcoming these difficulties, however, could contribute significantly to achieving quantum advantage. The Chicago Quantum Exchange (CQE) Pulse-level Quantum Control Workshop was a step toward progress in low-level quantum control. 

\bibliographystyle{unsrt}
\bibliography{refs}

\end{document}